\documentclass[twoside,11pt]{article}

\usepackage{tablefootnote}
\usepackage{jmlr2e}
\usepackage{amsmath}
\usepackage{amssymb,amsmath}
\usepackage{tikz}
\usepackage{listings}
\newcommand{\comment}[1]{}
\usepackage[utf8]{inputenc}
\usepackage{enumitem}
\usepackage{hyphenat}
\usepackage{color}
\usepackage{listings}
\usetikzlibrary{fit,positioning,arrows,automata}

\jmlrheading{1}{2022}{1-48}{4/00}{10/00}{Fernando Moreno-Pino,  Emese Sükei, Pablo M. Olmos, and Antonio Artés-Rodríguez}

\ShortHeadings{PyHHMM - Heterogeneous Hidden Markov Models}{Moreno-Pino, Sükei, M. Olmos, and Artés-Rodríguez}
\firstpageno{1}

\begin{document}

\title{PyHHMM: A Python Library for Heterogeneous Hidden Markov Models}

\author{\name Fernando Moreno-Pino \email fmoreno@tsc.uc3m.es\\
	\name Emese Sükei \email esukei@tsc.uc3m.es\\
	\name Pablo M. Olmos \email olmos@tsc.uc3m.es\\
	\name Antonio Artés-Rodríguez \email antonio@tsc.uc3m.es \\
	\addr Department of Signal Theory and Communications, Universidad Carlos III de Madrid, Spain
}

\maketitle

\begin{abstract}
We introduce PyHHMM, an object-oriented open-source Python implementation of Hete\-rogeneous-Hidden Markov Models (HHMMs). 
In addition to HMM's basic core functionalities, such as different initialization algorithms and classical observations models, i.e., continuous and multinoulli, PyHHMM distinctively emphasizes features not supported in similar available frameworks: a heterogeneous observation model, missing data inference, different model order selection criterias, and semi-supervised training. These characteristics result in a feature-rich implementation for researchers working with sequential data.
PyHHMM relies on the \textit{numpy, scipy, scikit-learn}, and \textit{seaborn} Python packages, and is distributed under the Apache-2.0 License. PyHHMM's source code is publicly available on Github\footnote{\url{https://github.com/fmorenopino/HeterogeneousHMM}} to facilitate adoptions and future contributions. A detailed documentation\footnote{\url{https://pyhhmm.readthedocs.io/en/latest}}, which covers examples of use and models' theoretical explanation, is available. The package can be installed through the Python Package Index (PyPI).
\end{abstract}

\begin{keywords}
Heterogeneous Hidden Markov Models, python, missing data inference, semi-supervised training
\end{keywords}

\section{Introduction}
\label{introduction}


Hidden Markov Models (HMMs), as defined by \cite{rabiner1989tutorial}, are generative models where the modeled system is assumed to be a Markov process, in which an observation model explains the observed data through a hidden variable.

Different frameworks that implement these well-known models are publicly available. However, these implementations usually lack features required to use these models with real-world datasets: missing data inference, ability to manage heterogeneity, semi-supervised training support to increase hidden states' interpretability, and synthetic data generation. Aiming to provide an HMM implementation that both academics and industry professionals can use, we present PyHHMM, an open-source Python toolbox that, in contrast to existing libraries, supports the previously enumerated features and constitutes a valuable alternative to work with sequential data.

\section{Heterogeneous Hidden Markov Model}
\label{section_hhmm}

Hidden Markov Models' objective is to learn the hidden states sequence, denoted $S=\left\{s_{1}, s_{2}, \ldots, s_{T}: s_{t} \in 1, \ldots, I\right\}$, with $t=\left\{1, 2, \ldots, T\right\} \in  \mathbb{N}$, that better explain the observed data, $Y=\left\{\mathbf{y}_{1}, \mathbf{y}_{2}, \ldots, \mathbf{y}_{T}: \mathbf{y}_{t} \in \mathbb{R}^{M}\right\}$. 
To do so, they use an observation model $p(\mathbf{y}_t | s_t)$. Therefore, each observation $\mathbf{y}_t$ depends exclusively on its associated state $s_t$.
 Nevertheless, standard HMM implementations use Multinomial and Gaussian observation models, depending on the probability distribution chosen to model the emission probabilities $p(\mathbf{y}_t | s_t)$.
 Heterogenous-HHMs (HHMMs) encompass these two va\-ria\-tions of the HMM, combining Gaussian and discrete observations. Therefore, the HHMM can manage heterogeneous data by using different observation emission probability distributions to model the conditional dependencies of the hidden states on the observations. 

\begin{figure}[!htb]
\centering
\begin{tikzpicture}
\tikzstyle{main}=[circle, minimum size = 10mm, thick, draw =black!80, node distance = 10mm]
\tikzstyle{connect}=[-latex, thick]
\tikzstyle{box}=[rectangle, draw=black!100]
  \node[main] (S1) [] {$s_{t-1}$};
  \node[main] (S2) [right=of S1] {$s_{t }$};
  \node[main] (S3) [right=of S2] {$s_{t+1}$};
  \node[main] (St) [right=of S3] {$s_{T_n}$};
  
  \node[main,fill=black!10] (O1) [below=of S1] {$y_{t-1}$};
  \node[main,fill=black!10] (O2) [right=of O1,below=of S2] {$y_{t}$};
  \node[main,fill=black!10] (O3) [right=of O2,below=of S3] {$y_{t+1}$};
  \node[main,fill=black!10] (Ot) [right=of O3,below=of St] {$y_{T_n}$};
  
  \node [main,fill=black!10] (L1) at (1,-1.2) {$l_{t-1}$};
  \node [main,fill=black!10] (L2) at (3,-1.2) {$l_{t}$};
  \node [main,fill=black!10] (L3) at (5,-1.2) {$l_{t+1}$};
  \node [main,fill=black!10] (Lt) at (7,-1.2) {$l_{T_n}$};

  \path (S3) -- node[auto=false]{\ldots} (St);
  \path (S1) edge [connect] (S2)
        (S2) edge [connect] (S3)
        (S3) -- node[auto=false]{\ldots} (St);

  \path (S1) edge [connect] (O1);
  \path (S2) edge [connect] (O2);
  \path (S3) edge [connect] (O3);
  \path (St) edge [connect] (Ot);
  \path (S1) edge [connect] (L1);
  \path (S2) edge [connect] (L2);
  \path (S3) edge [connect] (L3);
  \path (St) edge [connect] (Lt);

  \draw[dashed]  [below=of S1,above=of O1];
\end{tikzpicture}
\caption{Heterogeneous HMM architecture. Gray represents observed data.}
\label{hhmm}
\end{figure}
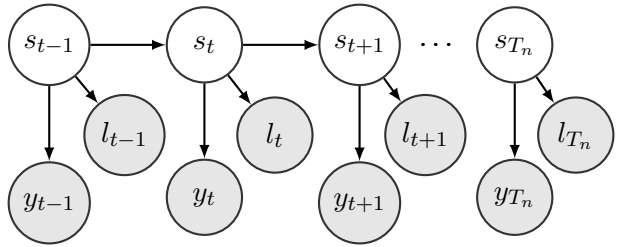

HHMMs can be fully characterized via the hidden states sequence, $S$; the continuous observations sequence, $Y$; its associated continuous observations emission probabilities, $\mathbf{B}=\left\{b_{i}: p_{b_{i}}\left(\mathbf{y}_{t}\right)=p\left(\mathbf{y}_{t} \mid s_{t}=i\right)\right\}$; the discrete sequence observations, $L=\{l_{1}, l_{2}, \ldots, l_{T}: l_{t} \in 1, \ldots, J\}$; its associated discrete observations emission probabilities, $\mathbf{D}=\{d_{i m}: d_{i m}=P\left(l_{t}=m \mid s_{t}=i\right)\}$; the state transition probabilities, $\mathbf{A}=\{a_{i j}: a_{i j}=p\left(s_{t+1}=j \mid s_{t}=i\right)\}$; and the initial state probability distribution, $\pi=\left\{\pi_{i}: \pi_{i}=p\left(s_{1}=i\right)\right\}$. HHMM's architecture is represented in Figure \ref{hhmm}.

 
\vspace{-0.25cm}

HHMMs, as classic HMMs, present three inference problems:
\begin{enumerate}[noitemsep, topsep=5pt]
    \item Given the observed data sequences $\{Y, L\}$ and the model's parameters $\theta = \{\boldsymbol{A}, \boldsymbol{B}, \boldsymbol{D}, {\pi}\}$, to estimate the probability of the observed sequence given the model, $p(Y, L | \theta)$.
    
    \item Given $\{Y, L\}$ and $\theta$, to determine the optimal hidden states sequence $S$ that better explains the data.
    This can be achieved by using the Forward-Backward algorithm, \cite{yu2003efficient}, calculating $p(s_t | \mathbf{y}_t, \mathbf{l}_t)$ each time-step, or through the Viterbi algorithm, \cite{forney1973viterbi}, which maximizes the probability of the hidden states sequences by considering all time-steps $t=\left\{1, 2, \ldots, T\right\}$, i.e., calculating $p(S | \mathbf{y}_t, \mathbf{l}_t)$. 

    \item To determine the optimal parameters $\theta$ that maximize the conditional probability $p(\mathbf{Y}, \mathbf{L} | \theta)$, which can be achieved through the Baum-Welch algorithm, \cite{frazzoli2013intro}. The joint distribution required for this third task, which may be modified to support any other type of observation emission probability' distribution, is expressed in Equation \ref{joint}.
\end{enumerate}

For more information regarding these problems and how to solve them, we refer the readers to \cite{rabiner1989tutorial}, \cite{anikeev2006parallel}, and \cite{cappe2009inference}.

\begin{equation}
\begin{aligned}
    p(S, Y, L)=\prod_{n=1}^{N}\left(p\left(s_{1}^{n}\right) \prod_{t=2}^{T_{n}} p\left(s_{t}^{n} | s_{t-1}^{n}\right)\right) \left(\prod_{t=1}^{T_{n}} p\left(\mathbf{y}_{t}^{n} | s_{t}^{n}\right)\right)\left(\prod_{t=1}^{T_{n}} p\left(l_{t}^{n} | s_{t}^{n}\right)\right)
\end{aligned}
\label{joint}
\end{equation}

\section{Library Implementation and Documentation}
\label{library_design}

 PyHHMM implements three different model's designs depending on the probability distribution that is chosen to manage the observed data: \textit{DiscreteHMM.py}, \textit{Gaussian\-HMM.py}, and \textit{HeterogenousHMM.py}. The three of them have a common parent class \textit{\_BaseHMM.py} that implements general functions for the three models, as the forward and backward methods, the log-probability computation, the training step, or the sampling functions.
 
 The library has dependencies on the Python packages \textit{numpy, scipy, scikit-learn}, and \textit{seaborn}. Documentation can be found on the Github repository as well as example notebooks that show simple use-cases to help new users understand the library’s functioning. These notebooks explain how to define the three available models, solve the described problems (training models, estimating data's likelihood, and decoding observation sequences), utilize the available model selection criteria, and sample data from the trained models. Theoretical properties of PyHHMM's implemented models and a description of the missing data inference process are also available.
 
 Regarding code quality, unit tests cover the implemented functions. The code is released under Apache-2.0  License and is available at \href{https://github.com/fmorenopino/HeterogeneousHMM}{Github}, which allows for collaborative development and facilitate the inclusion of prospect new features required by the community. Also, the package can be installed through PyPI \footnote{\texttt{pip install pyhhmm}}.

\section{Functionality and Comparison to other Software}
\label{functionality}

Three primary Python HMM implementations were identified: \textit{hmmlearn}\footnote{\url{https://hmmlearn.readthedocs.io/en/latest/index.html}}, \textit{pomegranate}\footnote{\url{https://pomegranate.readthedocs.io/en/latest/}}, and \textit{pyro}\footnote{\url{https://pyro.ai/examples/hmm.html}}.
While some of PyHHMM's features are also available in these libraries, none of these other pa\-cka\-ges support missing values inference, data heterogeneity, or semi-supervised training, core features while modeling real-world datasets. PyHHMM covers these necessities and some additional functionalities:

\begin{itemize}[noitemsep, topsep=2pt]
\item Missing data inference: available models can be trained in the presence of missing observations. Both cases, complete missing and partially missing, are covered by our implementation. Also, PyHHMM can perform missing value inference as a pre-processing technique. For further information regarding the inference process, consult \cite{murphy2012machine} and our package's documentation.

\item Semi-supervised training: 
PyHHMM allows the models to be trained in a semi-supervised manner, fixing discrete observations' emission probabilities, therefore using discrete sequences as labels. This semi-supervised training allows a guided learning process that improves the interpretability of the model parameters: certain states are associated with particular values of the labels.

\item Model selection criteria: Akaike Information Criterion (AIC) and the Bayesian Information Criterion (BIC) are implemented to estimate the model’s optimal number of hidden states.

\item Diagonal, full, tied, or spherical covariance matrices can model the Gaussian observations' emission probabilities.


\item Synthetic data generation, sampling from the trained models.

\end{itemize}
\comment{

Plenty of implementations of the HMM are available online. Three main Python pa\-cka\-ges on HMMs were identified: \textit{hmmlearn}\footnote{\url{https://hmmlearn.readthedocs.io/en/latest/index.html}}, \textit{pomegranate}\footnote{\url{https://pomegranate.readthedocs.io/en/latest/}} and \textit{pyro}\footnote{\url{https://pyro.ai/examples/hmm.html}}. Other libraries are also available in different languages, as \textit{ghmm}\footnote{\url{http://ghmm.org}}, implemented in C, or \textit{Matlab's implementation}\footnote{\url{https://es.mathworks.com/help/stats/hidden-markov-models-hmm.html}}.
While some of the previously enumerated features of PyHHMM are also available in some of these libraries, as the capability of generating samples from trained models or supporting different types of covariances matrices for gaussian observations, none of these other pa\-cka\-ges implement missing value support, heterogeneous models neither semi-supervised training. These features can be crucial while modeling real-world data and that is the necessity that our library covers.
}


\section{Basic Example}
\label{example}

This section provides a basic example of training a Heterogeneous-HMM over a set of pre-loaded sequences and using the resulting model to perform decoding. Examples of obtaining the AIC of the trained model and how to generate new synthetic samples using the resulting architecture are also included.


\begin{lstlisting}[language=Python, basicstyle=\small]
>>> from pyhhmm.heterogeneous import HeterogeneousHMM
>>> import pyhhmm.utils as ut

>>> my_hhmm = HeterogeneousHMM( n_states=2, n_g_emissions=2, 
              n_d_emissions=2, n_d_features=[2, 2], init_type="random", 
              covariance_type="diagonal", nr_no_train_de=1, 
              verbose=True ) # Initializing a HeterogeneousHMM
          
>>> my_hmm, log_likelihood = my_hhmm.train( training_seq, n_init=1, 
                             n_iter=50, thres=0.001, conv_iter=5,
                             plot_log_likelihood=True ) # Training

>>> logL, state_seq = my_hhmm.decode([training_seq[0], 
                      algorithm="viterbi") # Decoding
                      

>>> dof =  ut.get_n_fit_scalars(temp_ghmm) #Free parameters of the model
>>> AIC = ut.aic_hmm(log_likelihood,  dof) # AIC of the trained model

# Generating new sampled data with the trainned model
>>> X, state_sequences = my_hhmm.sample(n_sequences=n_sequences, 
                         n_samples=n_samples)
\end{lstlisting}

\section{Summary}
\label{conclusion}

PyHHMM is a flexible toolbox currently being used in academia and industry projects. By implementing well-known HMMs algorithms, this library provides a proper solution to several problems, such as obtaining quick predictors for discrete-time stationary processes, generating new synthetic data, or performing missing data inference. Detailed documentation and several examples are available on the project's site.

\acks{}

This work has been supported by the Spanish government Ministerio de Ciencia, Innovación y Universidades under grants FPU18/00470, TEC2017-92552-EXP and RTI2018-099655-B-100, by Comunidad de Madrid under grants IND2017/TIC-7618, IND2018/TIC-9649, IND2020/TIC-17372,  and Y2018/TCS-4705, by BBVA Foundation under the Deep-DARWiN project, and by the European Union (FEDER) and the European Research Council (ERC) through the European Union’s Horizon 2020 research and innovation program under Grant 714161 and Marie Sklodowska-Curie grant agreement No 813533.

\vskip 0.2in

\end{document}